\documentclass[aps,prl,amsmath]{revtex4}
\usepackage{graphicx}
\usepackage{color}
\usepackage{bm}
\usepackage{epsfig}
\usepackage{latexsym}
\usepackage{pifont}
\usepackage{float}
\usepackage[utf8]{inputenc}
\graphicspath{{figure/}}
\usepackage{siunitx}
\usepackage{textgreek}
\usepackage{amssymb}
\usepackage{appendix}
\begin{document}

\newcommand{\be}{\begin{equation}}
\newcommand{\ee}{\end{equation}}
\newcommand{\bea}{\begin{eqnarray}}
\newcommand{\eea}{\end{eqnarray}}
\newcommand{\nn}{\nonumber}
\newcommand{\ra}[1]{\textcolor{red}{#1}}
\newcommand{\ba}[1]{\textcolor{blue}{#1}}
\title{The Motile-Units model: Interacting spins model of cell polarization and motility}

\author{Jonathan E. Ron$^{1}$, Nir S. Gov$^{1}$}
\affiliation{$^{1}$ Department of Chemical and Biological Physics, Weizmann Institute of Science, Israel}

\begin{abstract}
We introduce a coarse-grained interacting-spin model for two-dimensional cell motility, in which the cell perimeter is discretized into stochastic binary spins that switch between active and inactive states. Each perimeter spin represents a "motile-unit" that is a source of protrusive force and retrograde flow when active. Long-range interactions between the motile-units arise through a polarity cue advected by the collective actin retrograde flow, providing a minimal realization of spontaneous symmetry breaking and self-propulsion. The model exhibits three dynamical phases, a random walk phase, persistent random walk phase, and an intermittent bistable phase characterized by run-and-tumble migration. Additional nearest-neighbor interactions modulate speed and persistence without altering the overall phase structure. Owing to its simplicity, the framework naturally incorporates external cues, reproducing chemotactic migration, steering by localized optogenetic activation, and directional decision-making (symmetry breaking) under competing stimuli. The model introduces a new class of active-particle model in which both speed and polarity emerge from internal stochastic spin dynamics, rather than being imposed as particle-level variables, offering a framework for the study of cell migration and extends the scope of active-matter physics. 
\end{abstract}

\maketitle

\section{Introduction} \label{section:MU}
Theoretical models of motile cells range in their level of detail and sophistication \cite{mogilner2009mathematics,holmes2012comparison,aranson2016physical,saez2026computational}. The most basic trait that models of motile cells possess is a description of a mechanism that can lead to spontaneous symmetry breaking and polarization of the cellular cytoskeleton, such that the cell can persistently exert traction forces that propel it on some substrate. The biological complexity of the cytoskeleton is huge, with hundreds of proteins involved in the motility process. Theoretical models therefore reduce this complexity to just a few components, each model focusing on a slightly different component set. The challenge is to have models that contain enough components to allow for the emergence of polarity and motility, while keeping this number to a minimal value to allow the model to be comprehensible and amenable to analysis, as well as remaining simple to use and computationally efficient.

There are numerous examples of such models, based on different mechanisms, such as biochemical signaling networks \cite{iglesias2026signals}, or adhesion-driven feedback \cite{sens2020stick}. Here we wish to explore a variation of a model where the front-back polarization of the cell is driven by a positive feedback between the activity of the acto-myosin cytoskeleton at the cell periphery and the advection of different cellular components by the resulting net actin treadmilling flow. This model was developed in one-dimension \cite{maiuri2015actin}, and termed universal coupling between speed and persistence (UCSP). It was later expanded to describe more complex migration patterns in one-dimension \cite{lavi2016deterministic,ron2020one,mukherjee2023actin}, on networks \cite{ron2024emergent,liu2026trade,liu2026modelling}, and for one-dimensional multi-cellular "trains" \cite{ron2023polarization}. 

Generalization of this model to a two-dimensional cell model \cite{lavi2020motility} become highly complicated when attempting to solve for the full cytoplasmic flow and cell contour dynamics. Simplifying to a circular cell of rigid shape, while solving the full intracellular flows and resultant actin polymerization profile, is also highly demanding \cite{etchegaray20252d}.

Here we propose a simplified approach, conceptually motivated by our spin-based model developed to describe collective food transport by ants \cite{feinerman2018physics,ron2018bi}. In our model (Fig.\ref{fig:MU_01}A), the cell is represented as a rigid disk whose rim is divided into $N$ "Motile Units" (MU), each with a binary "on"/"off" degree of freedom that is described as a "1"/"0" spin. When the MU is "on" (spin value "1") it exerts a protrusion force directed at the local outwards normal to the cell rim. The transition rate at which an MU spin flips from one state to the other depends on the local distribution of an inhibitory cue along the rim, driven by the retrograde flow. In order to account for the UCSP mechanism, we simplify the two-dimensional problem by assuming that the polarity cue forms a one-dimensional concentration gradient along the net retrograde flow (polarization) axis, which is projected onto the cell rim. Since in our model the UCSP polarity cue acts as an inhibitor of acto-myosin activity \cite{ron2020one,mukherjee2023actin,ron2024emergent,liu2026modelling,liu2026trade}, its role in the spin model is to increase the "on"$\rightarrow$"off" spin flip rate. 


A key motivation behind dividing the cell rim into discrete force-generating elements (Motile Units) is that protrusion formation along the cell periphery drive migration. These protrusions span a wide range of spatial scales, which maens that depending on the level of coarse-graining, a MU may represent different protrusive structures, such as local actin rich protrusive domains ($\sim0.1$--$1~\mu$m) \cite{pollard2003cellular}, filopods ($\sim1$--$10~\mu$m) \cite{svitkina2003mechanism}, blebs ($\sim1$--$10~\mu$m) \cite{charras2008blebs,lavi2019cellular}, and pseudopods ($\sim5$--$20~\mu$m) \cite{insall2013interaction}. Moreover, the organization and dynamics of protrusions along the cell periphery determine the mode of migration \cite{insall2010understanding,matsumoto2026statistical}. Cells are often found to spontaneously transition between amoeboid and lamellipodial migration modes, where amoeboid migration is characterized by a tug-of-war between multiple protrusions in different directions and low directional persistence, whereas lamellipodial migration is dominated by a single polarized leading edge \cite{moldenhawer2022spontaneous}.

The role of these distinct cellular protrusions in propelling cell migration has motivated several theoretical and computational models. Some models rely on reaction-diffusion dynamics that spontaneously splits the cellular protrusion as the cell migrates \cite{neilson2011modeling}, utilizing phase-field models to describe complex two-dimensional cell shapes \cite{niculescu2015crawling,burger2022density}, and describing protrusions undergoing regular cycles of growth and retraction \cite{allena2013cell,heck2020role}. A recent extension of these models also includes the effects of flows in the cytoplasm \cite{ivvsic2026diversity}. Another model that discretizes the cell rim into force-producing units relies on reaction-diffusion to provide global polarization \cite{merchant2018rho}. Model of cell migration based on the Cellular Potts Model (CPM) are also naturally discretizing the protrusive activity along the cell rim \cite{wortel2021local}. In \cite{etchegaray2018stochastic} the cell polarity is driven by an unspecified feedback between cell speed and activation of cellular protrusions, which we explicitly describe using the UCSP mechanism in our model.

In our proposed model, we offer a simple coarse-grained approach to predict cell migration and cell polarization. In this approach, competition between protrusive activity along the cell periphery is the dominant mechanism, while the shape dynamics are not explicitly described. This simplification reduces the computational effort required for simulations, while retaining the essential mechanisms underlying cell polarization and migration. Recent models in the spirit of our proposed approach include model of pseudopod-based chemotaxis \cite{alonso2025persistent}, where the cell polarity regularly splits into $N$ competing protrusions, which affect each other through the depletion of a common resource (actin monomers). Another protrusion-based model has microtubule-driven positive feedback as the polarization mechanism \cite{vaidvziulyte2022persistent}. Several recent models have also explored cell migration through random protrusions, lacking direct competition and focusing on the role of adhesion \cite{louviaux2026mechanics,ALLENA2026112550}. 

In contrast to these models, our model offers a different global competition mechanism between protrusive activity along the cell rim, mediated by the UCSP mechanism \cite{ron2020one,ron2023polarization}. The spin description allows us to naturally include stochasticity through the "on"/"off" spin flip dynamics. We demonstrate the applicability of this new theoretical framework to describe chemotactic migration and directional steering by photoactivation \cite{town2023local}. The model also expands the scope of active-matter physics by introducing a new type of active, self-propelled particle, that goes beyond the treatment in terms of particle-level variables: here both speed and polarity emerge from internal stochastic spin dynamics, and even exhibits rudimentary directional decision making abilities \cite{gompper20252025}.



\section{The MU Model}
\begin{figure} [htbp] 
\centerline{\includegraphics[width=1\textwidth]{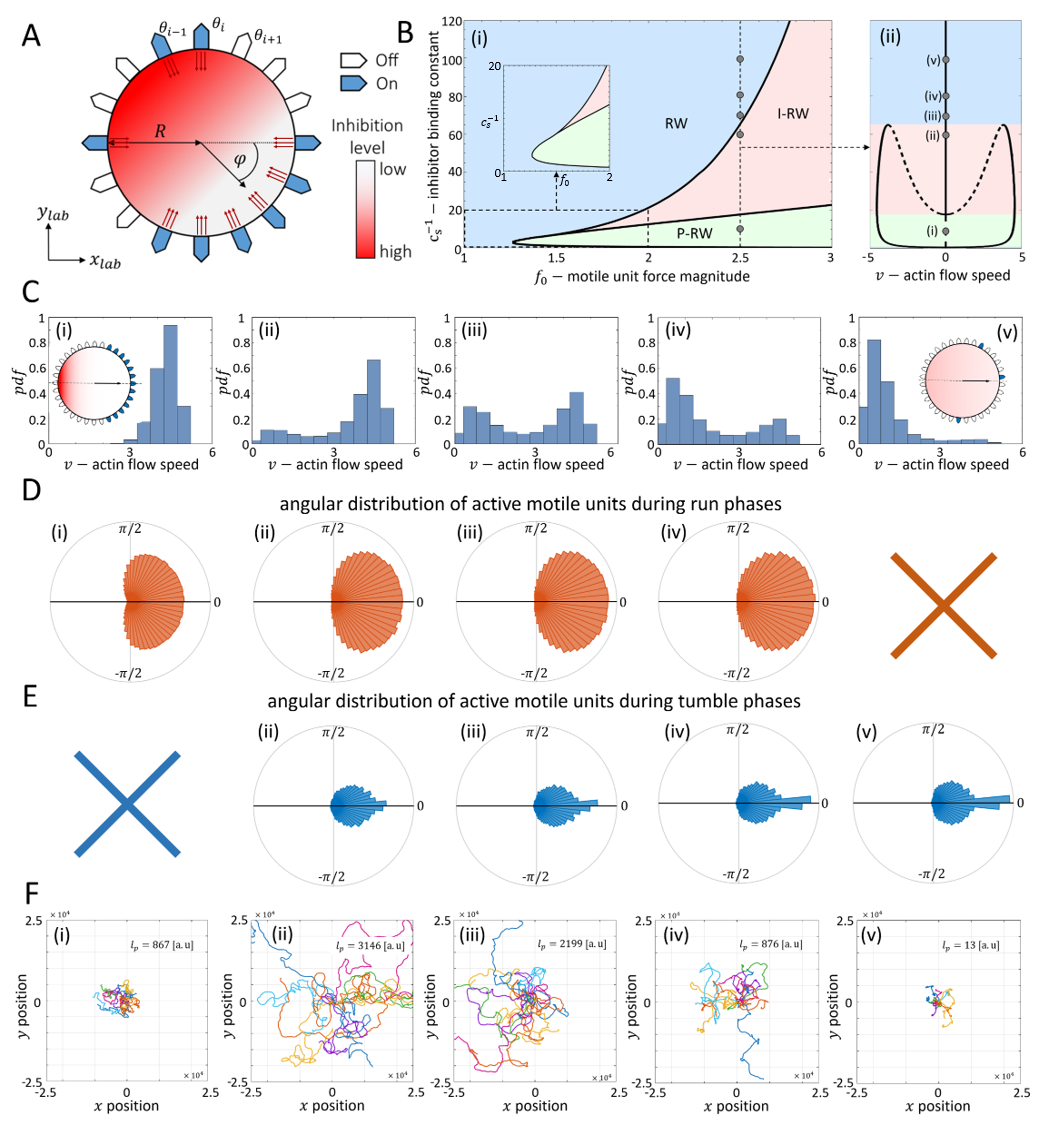}}
\caption{\footnotesize{A) Schematic of the 2D motile-units model. Blue and white trapezoids denote active and passive motile units, respectively. The red-white gradient indicates the concentration of the polarity cue that inhibits unit activity, given by Eq.(\ref{eq:c}). $R$ is the cell radius, $\theta_i$ is the angular position of unit $i$ along the rim, and $\varphi$ is the polarization axis relative to the $x$-axis. The polarization axis is along the direction of the total force. The retrograde flow contribution from each active MU is denoted by the red arrows. B) (i) Phase diagram of the $c_s^{-1}$-$f_0$ parameter space. The blue, red, and green regions correspond to the random-walk (RW), intermittent random-walk (I-RW), and persistent random-walk (P-RW) regimes, respectively. (ii) The bifurcation profile of the steady-state actin flow speed $v$ as function of $c_s^{-1}$, along a vertical cross section at $f_0=2.5$ (vertical dashed line in (i)). Solid lines denote the stable solutions and dashed lines the unstable solutions (in the I-RW phase). The inset enlarges the region where the three phases meet, to show the RW phase at very low values of $c_s^{-1}$. Gray points indicate the parameter values used in panels (C,D): $c_s^{-1}=10,60,70,80,100$. C) Probability density functions of the actin flow speed $v$ for $N=32$ units at the parameter values marked in (B). Insets in (i) and (v) show representative instantaneous configurations of the active MU around the cell.
(D-E) Angular distributions of active MUs during run (red) and tumble (blue) phases, measured relative to the instantaneous direction of motion ($\varphi$), for $c_s^{-1}=10,60,70,80,100$. The run–tumble threshold is determined as the local minimum in the velocity distributions (C), between the low and high speed peaks. Each polar histogram is normalized by the total number of active-MU such that the sum over all angular bins is unity. The symbols $X$ denotes the lack of tumble or run phases in these parameter values. F) Examples of cell trajectories for the same parameter values as in (C). The persistence length (decay length of the velocity-velocity correlation function) $l_p$ is given in each panel. Other model parameters: $R=1$,$D=1$,$c_{tot}=1$.}}
\label{fig:MU_01}
\end{figure}

We model a cell migrating in a two dimensional environment as a rigid disk of radius $R$. The cell rim is discretized into $N$ equally spaced Motile Units (MUs), located at angular positions $\theta_i$ (Fig.\ref{fig:MU_01}A). Each MU can be active or passive. Active MU exert an outward radial protrusive force of magnitude $f_0$, whereas passive MU exert no force (blue and white outwards arrows along the cell rim in Fig.\ref{fig:MU_01}A).

The average probability that a MU is active obeys a linear kinetic model
\begin{equation} \label{n_dot}
\dot{n}\left(\theta_i\right)=k_{on}(1-n\left(\theta_i\right))-k_{off}n\left(\theta_i\right)c\left(\theta_i\right)
\end{equation}
where $k_{on}$ and $k_{off}$ are the basal activation and deactivation rates, respectively. The activation rate is constant, whereas the deactivation rate is proportional to the local concentration of the inhibitory polarity cue
along the rim $c(\theta_i)$. At steady state ($\dot{n}(\theta_i)=0$), the probability that a MU is active is given by
\begin{equation}
n\left(\theta_i\right)=\frac{c_s}{c_s+c\left(\theta_i\right)}
\label{ntheta}
\end{equation}
where $c_s=k_{on}/k_{off}$. Throughout this work we analyze the system using the inverse parameter $c_s^{-1}$, which we define as the inhibitor binding constant.

The local inhibitor concentration at each MU along the rim, $c(\theta_i)$, is obtained by projecting the one-dimensional concentration gradient along the polarization axis along the cell rim (Eq.\ref{c2D})
\begin{equation} \label{eq:c}
     c(\theta_i) = \frac{c_{tot} v}{2\pi R D}\frac{exp\left(-\frac{v R}{D}cos(\theta_i-\varphi)\right)}{I_1\left(\frac{v R}{D}\right)}
\end{equation}
where $\varphi$ is the angle of the polarization axis (direction of the total force, Eq.(\ref{vss2D})) with respect to the $x$-axis of the lab frame (Fig.\ref{fig:MU_01}A).
where $I_1$ is the Bessel function of the first kind, and $v$ is the magnitude of the retrograde flow which will be described below. The derivation of the concentration profile is given in Appendix A, and is based on the assumption that the concentration profile reaches its steady-state shape, given by the balance between advection and diffusion while maintaining overall conservation of the total amount of inhibitor in the cell (as was previously assumed in \cite{maiuri2015actin,ron2020one}).

The assumption that the inhibitor gradient is only along the one-dimensional axis of the polarization simplifies the analysis, which would otherwise involve solving complex velocity fields \cite{lavi2020motility,etchegaray20252d}. This is a minimal way of implementing long-range interaction that is based on the advection-diffusion of the inhibitory cue that is sensed at the rim of the cell.

The stochastic simulations are performed using the discrete MU model, where each MU occupies one of two states, active ($n(\theta_i)=1$) or passive ($n(\theta_i)=0$). The total protrusive force exerted by the cell is therefore a sum over the active forces (blue arrows in Fig.\ref{fig:MU_01}A)
\begin{equation} \label{vss2D}
    \vec{f} = f_0\frac{2\pi}{N}\sum_{i=1}^{N} n(\theta_i)\left(cos(\theta_i)\hat{x}+sin(\theta_i)\hat{y}\right)
\end{equation} 
The factor of $2\pi/N$ accounts for the fact that the relative size of the MU as a fraction of the cell rim gets smaller as their number increases.

The same active MUs generate an actin retrograde flow $\vec{v}$, that is directed opposite to the total protrusive force (red arrows from each active MU in Fig.\ref{fig:MU_01}A). For simplicity, we assume that the net retrograde flow is linearly proportional to the total protrusive force,
\begin{equation}
    \label{vret}
    \vec{v} = -\mu \vec{f}
\end{equation}
where $\mu$ is the proportionality coefficient. Throughout this work, we choose units such that $\mu=1$.

Using Eq.(\ref{vss2D}), the retrograde flow is therefore given by
\begin{equation}
    \label{vret_discrete}
    \vec{v}
    = -f_0\frac{2\pi}{N}\sum_{i=1}^{N} n(\theta_i)
    \left[
    cos(\theta_i)\hat{x}
    +sin(\theta_i)\hat{y}
    \right]
\end{equation}
Thus, in the units used here, the total protrusive force and the retrograde flow have the same magnitude but point in opposite directions.

To convert the total protrusive force into the cell migration velocity, we assume an overdamped force--velocity relation,
\begin{equation}
    \label{vcell}
    \vec{V}_{\rm cell}=\gamma\vec{f}
\end{equation}
where $\gamma$ is the inverse of an effective friction coefficient. We choose units such that $\gamma=1$. A more detailed description of the cell--substrate mechanical coupling could allow this effective friction to depend on properties such as the length of the cell rim \cite{ron2026theory}, and could additionally account for variations in cell-substrate adhesion associated with the activity states of the MUs. In general, the parameters $\mu$ and $\gamma$ are related to the strength of the actin-adhesion (clutch-like) coupling \cite{bangasser2013determinants,alonso2023optimal}, and can be implemented in more complex (and realistic) forms, such as biphasic dependence between the adhesion and retrograde flow \cite{jurado2005slipping,gardel2008traction}.

Consequently, in the units used throughout this work,
\begin{equation}
    \vec{V}_{\rm cell}=\vec{f}=-\vec{v}
\end{equation}
so that the protrusive force, cell speed, and retrograde-flow speed have the same numerical magnitude, while the cell velocity and retrograde flow point in opposite directions.

The transition rate dynamics of the activation and deactivation of the MU (spin flipping events) are calculated using a stochastic Gillespie scheme \cite{gillespie1976general,feinerman2018physics,ron2018bi}. For a passive (active) MU at position $\theta_i$, the transition rate to the active (passive) state is given by $k_{on}$ ($k_{off}c(\theta_i)$). These  are the same kinetic rates introduced in Eq.(\ref{ntheta}).

During the waiting time $\delta t$ between two consecutive spin-flip events, the MU configuration and hence the cell velocity are taken to remain constant. The cell position is therefore updated according to
\begin{equation}
    \label{cell_position}
    \vec{r}(t+\delta t)
    =
    \vec{r}(t)
    +
    \vec{V}_{\rm cell}(t)\delta t
\end{equation}
Thus, the stochastic MU dynamics determine the instantaneous polarization and migration velocity of the cell, while the cell trajectory is obtained directly from the kinematic integration of this velocity over the sequence of Gillespie waiting times.

Note that in our model the stochasticity arises explicitly due to the spin-flip dynamics that turn MU on/off according to the  kinetic rates. The spin-flip induced variance in the total force and speed of the cell decreases as $1/N$ \cite{pinkoviezky2018collective}. In our calculations there are no additional noise sources beside this intrinsic source.


Since only radial protrusive forces are considered in the model, no torque is generated and the cell does not rotate. Similar to our spin-based model of cooperative transport by ants \cite{feinerman2018physics,ron2018bi}, future extensions of this model may include MU forces with non-radial components, which can generate rotations.

\section{Migration modes of the MU model}

For the analytic analysis of the model, we set the $x$-axis as the axis of polarization and use the continuum limit of Eq.(\ref{vret_discrete}) for the retrograde flow speed $v$. Note that the cell speed is simply opposite to the direction of the retrograde flow. We define the functional $F(v)$ whose zeros give the mean-field (MF) solutions 
\begin{equation} \label{v_ss}
v=f_0\int_{0}^{2\pi}n(\theta)cos(\theta)d\theta\longrightarrow F(v) = -v+f_0\int_{0}^{2\pi}n(\theta)cos(\theta)d\theta
\end{equation}

We choose to keep the dimensional form of Eq.(\ref{v_ss}) without rescaling to allow us to independently examine the role of each parameter which might be in principle amenable to experimental manipulation. Yet, we note that it is natural to rescale the system by the dimensionless Pecl\'et number ($v R/D$)  \cite{ron2026theory} and the inhibitor concentration by the ratio ($c_{tot}/c_s$) \cite{ron2020one} (the radius was fixed at $R=1$).

In Fig.\ref{fig:MU_01}B(i) we display the $f_0-c_s$ phase diagram based on the analysis of $F(v)$ (Eq.(\ref{v_ss})), which provide the mean-field (MF) behavior. The results show that the model predicts three different motility regimes, similar to the behavior in the 1D model \cite{maiuri2015actin,mukherjee2023actin}: 1) Random walk regime (RW) - where $F(v)$ has a stable solution at $v=0$, and the cell exhibits a Brownian motion (Fig.\ref{fig:MU_01}B(ii)). 2) Persistent random walk regime (P-RW) - where $F(v)$ has a stable non-zero solution, and the cell exhibits a persistent polarized motion (Fig.\ref{fig:MU_01}B(ii)). 3) Intermittent random walk regime (I-RW), where $F(v)$ has a stable zero and non-zero $v$ solutions, and the cells display bi-stability between random walks and persistent random walks (Fig.\ref{fig:MU_01}B(ii)).

In Fig.\ref{fig:MU_01}C we plot the velocity probability density functions (PDFs) obtained from simulations at the five parameter values marked by the gray circles in Fig.\ref{fig:MU_01}B. The lowest point (i), in the P-RW regime, exhibits a single finite-velocity peak, in good agreement with the mean-field (MF) prediction shown in Fig.\ref{fig:MU_01}B(ii). Fig.\ref{fig:MU_01}C(ii--iv) illustrates the gradual crossover from the I-RW to the RW regime As $c_s^{-1}$ increases. At point (ii), the velocity distribution exhibits the expected double peak of the I-RW regime. At points (iii) and (iv), the MF theory predicts a transition to the RW regime, with a single peak at zero velocity. However, due to the finite-size simulations the peak at a non-zero velocity remains clearly visible even inside the RW regime predicted by the MF phase diagram. Finally, Fig.\ref{fig:MU_01}C(v) exhibits the single peak at zero velocity expected for the RW regime, indicating that the finite-size effects have become negligible.

Note that while the MF phase-diagram (Fig.\ref{fig:MU_01}B) is independent of the number of MU, $N$, its strictly giving the transitions in the limit of $N\rightarrow\infty$. However, clearly the velocity distributions shown in Fig.\ref{fig:MU_01}C depend on $N$ (See Appendix B). As $N$ increases (decreases) these distributions have smaller (larger) variance \cite{pinkoviezky2018collective}, while the locations of the peak remain roughly unchanged (corresponding to the MF solutions). A similar effect was shown in another protrusion-based cell migration model \cite{heck2020role}, where a smaller number of protrusions corresponds to more persistent cellular migration. 

\begin{figure} [htbp] 
\centerline{\includegraphics[width=1\textwidth]{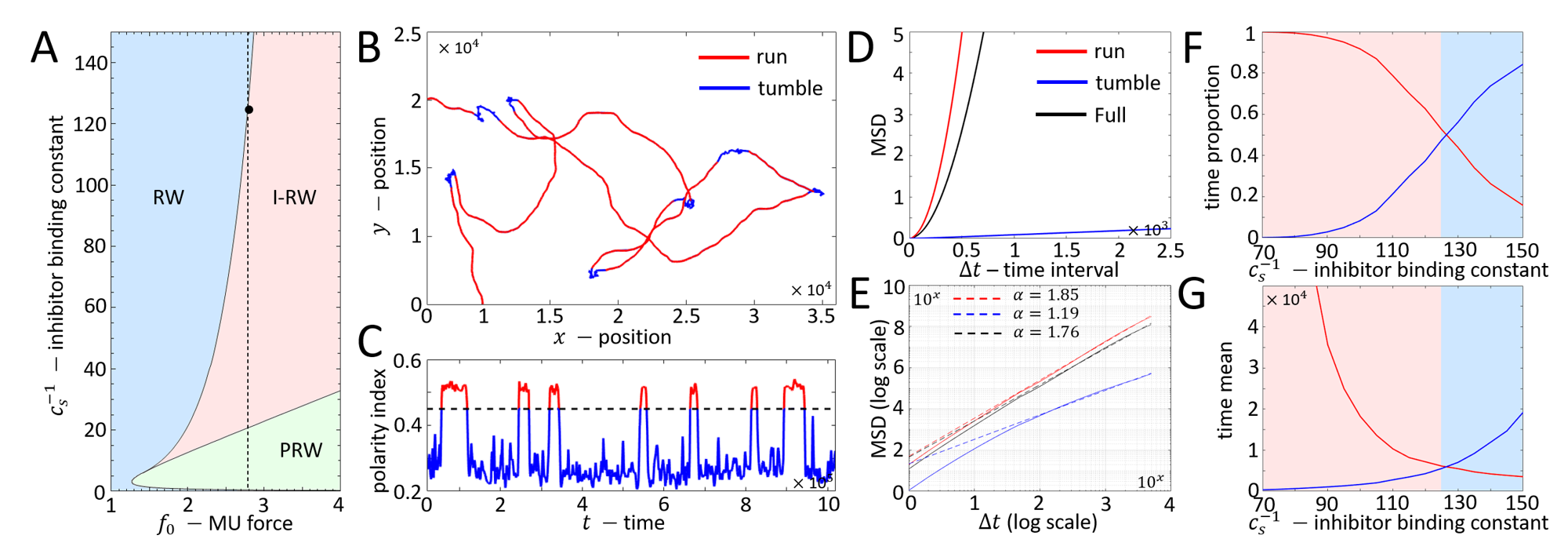}}
\caption{\footnotesize{ A) The $c_s^{-1}-f_0$ parameter space. Dashed line indicate the cross-section of $f_0=2.8$. Black Point is $(c_s^{-1},f_0)=125,2.8$. B) Cell trajectory. Red and blue indicate the run and tumble phases respectively. C) The polarity index as a function of time. Dashed horizontal line indicates the threshold used to separate between the run (red) and tumble (blue) phases. D) The mean-squared-displacement (MSD) of the tumble phase (blue), run phase (red) and the full trajectory (black). E) Log scale of the mean squared displacement (Solid line, linear fit - dashed line). $\alpha$ is the diffusion exponent. F) The proportion of the time spent in each phase (tumble - blue, run - red) along the cross-section of $f_0=2.8$ in A. G) The average time spent in each phase (tumble - blue, run - red) along the cross-section of $f_0=2.8$ in A. Other model parameters: $R=1$,$D=1$,$c_{tot}=1$.}}
\label{fig:MU_02}
\end{figure}

In addition, in the I-RW regime there are two peaks in the distribution (Fig.\ref{fig:MU_01}C), and the transitions between them (Fig.\ref{fig:MU_02}) occur more (less) frequently as $N$ is decreased (increases) \cite{pinkoviezky2018collective,ayalon2021sequential}. We chose here the value of $N=32$, which is a compromise which gives high angular resolution for the 2D motion, while maintaining the computations efficient and fast. From the perspective of the cell, the realistic value of $N$ may depend on the typical size of the protrusions (lamellipodia, invadopodia, filopodia or blebs) along the cell edge \cite{pollard2003cellular,svitkina2003mechanism,charras2008blebs,lavi2019cellular,insall2013interaction}. In most cells these protrusions are much larger than single actin filaments, and typically $N\sim O(1-10)$ \cite{neilson2011modeling,vaidvziulyte2022persistent}. 

As shown in Fig.\ref{fig:MU_finite_size} (Appendix B), when the number of MU is smaller, whereby each MU covers a relatively larger part of the cell edge, the migration properties become more persistent. This means that when fitting experimental cell migration data to the MU model, the appropriate number of MU ($N$) will affect the fitted values of the model parameters ($f_0,c_s^{-1}$). Celllular protrusions by blebs, for example, may correspond to relatively small $N$, with blebs typically having a regular size that covers a significant portion of the cell surface \cite{diz2016steering,olguin2021chemokine}.

In Fig.\ref{fig:MU_01}D,C we show the angular distribution of the active MU for cells with the parameters denoted by points (i-v) in Fig.\ref{fig:MU_01}B. We denote the "run" and "tumble" phases by the instantaneous velocity of the cell: velocities that are below the minimum in the velocity distributions (Fig.\ref{fig:MU_01}B(ii-iv)), are associated to the tumble phase, while those above it are associated with the run phase. In the PRW (point (i)) there is no tumble phase, and the active MU form a large leading-edge cluster. As $k_{on}$ decreases ($c_s^{-1}$ increases), the leading-edge cluster of active MU decreases in its angular spread 
(Fig.\ref{fig:MU_01}D(ii-iv)).

In Fig.\ref{fig:MU_01}F we show typical simulation trajectories for the parameter values of Fig.\ref{fig:MU_01}C-E (gray points indicated in Fig.\ref{fig:MU_01}B). In the P-RW regime (point (i)), the cell has a relatively large average speed, however the directional persistence is moderate, as measured by the velocity-velocity directional correlations (see Appendix C, Fig.\ref{fig:MU_app_D}). For lower $c_s^{-1}$ in the I-RW regime (point (ii)) we expect the directional persistence to decrease due to the tumbling events, which appear as the peak around zero velocity (Fig.\ref{fig:MU_01}C(ii)). However, we find that the persistence length of the cellular directional migration is greatly increased when compared to point (i). This is an unexpected trend, whereby the directional persistence increases despite a decrease in the average speed (as can be seen in the velocity distribution), in contrast to the one-dimensional UCSP model \cite{maiuri2015actin,ron2020one,mukherjee2023actin}. 

We can explain this behavior as follows: In the P-RW regime $c_s^{-1}$ is relatively small ($k_{on}$ is large), and the size of the leading edge cluster is very large (Fig.\ref{fig:MU_01}C(i),D(i)). The edges of this cluster extend to large angles on either side of the polarity axis, such that on/off fluctuations of MU induce sizable cell velocity fluctuations that are orthogonal to the polarization axis, thereby introducing frequent and large directional re-orientations of the polarization (and migration) axis. A calculation based on these orthogonal velocity fluctuations provides an estimation of the persistence time and length, as shown in Appendix C.

As $c_s^{-1}$ increases (point (ii) in Fig.\ref{fig:MU_01}B-D), $k_{on}$ decreases, and the leading edge cluster decreases in size, becoming more strongly confined to the front of the cell. In such a well-confined leading edge cluster there are only very small angle fluctuations to the retrograde current, and therefore a lower rate at which the migration direction changes.

As we move deeper into the RW regime (increasing $c_s^{-1}$), both the average velocity and the persistence length decrease (Fig.\ref{fig:MU_01}D), as expected in the UCSP model \cite{maiuri2015actin}. This is due to the leading edge cluster losing its stability, as the gradient of the inhibitory polarity cue across the cell diminishes and the overall number of active MU diminishes (see typical snapshot of the MU states in Fig.\ref{fig:MU_01}C(v)).

In Fig.\ref{fig:MU_02} we study further the run-and-tumble behavior exhibited in the I-RW regime. We choose the point $\left(f_0,c_s^{-1}\right)=(2.8,125)$ which is located near the transition between the RW and I-RW regimes (Fig.\ref{fig:MU_02}A). We choose this point since we found that due to the final size effect we can present clear visualization of the the trajectories and their analysis. In Fig.\ref{fig:MU_02}B we show a typical trajectory which distinguishes between the run (red) and the tumble (blue) events, and analyze it similar to the analysis shown in \cite{nathan2022big}.

The classification between the runs and the tumbles (Fig.\ref{fig:MU_02}C) is done using a polarity index we defined by
\begin{equation}
p=\frac{1}{N}\sum\limits_{i=1}^{N}|(1+9\,n_i(\theta_i))|cos(\theta_i-\varphi)
\label{polarity}
\end{equation}
where $|n_i(\theta)|=0,1$ with respect to the state of the motile unit. The threshold of the polarity index is set to $p=0.5$. The motivation for this definition is explained in Appendix D.

The properties of the run-and-tumble trajectories are shown in Fig.\ref{fig:MU_02}D,E where the mean-square-displacement (MSD) is plotted as function of time, for the whole trajectory and for the run and tumble sections separately. We find that the tumble phase part of the trajectory is RW-like with an MSD exponent $\alpha$ ($\langle r^2 \rangle\propto t^{\alpha}$) close to $1$. In the run phase, this exponent is close to $2$, indicating that for the parameters shown here the run phase has a directional persistence time that is longer than the times between consecutive tumble events. As function of the $c_s^{-1}$ control parameter (along the vertical dashed line shown in Fig.\ref{fig:MU_02}A) we plot in Fig.\ref{fig:MU_02}F the average proportion of the time spent in the two phases (same colors as in B). The average time durations spent in the two phases are shown in Fig.\ref{fig:MU_02}G. We find that closer to the P-RW phase the run phase dominates, while closer to the RW phase the tumble phase increases in duration such that the two phases are equally dominant at the RW transition.


\section{Effects of nearest-neighbor MU interactions}

We now extend the basic MU model to demonstrate how it can incorporate additional direct interactions between the MU. Experiments suggest that there are local positive feedback mechanisms between neighboring regions of activated cellular protrusions at the cell rim \cite{weiner2002ptdinsp3,inoue2008synthetic,lebensohn2009activation,wu2025wave}. This local positive feedback, combined with long-range inhibition, is the basis of many models of polarized cell migration \cite{xiong2010cells,iglesias2026signals}. 

In our model, this direct nearest-neighbor (NN) positive feedback is not essential for the polarization of the cell, as we showed above. Nevertheless, we show below that local interactions between neighboring MU modify the dynamics of the cell.

To incorporate NN interaction in the stochastic simulations, each MU is described by a binary state
\begin{equation}
s_i = \begin{cases} 
      1, & \text{"off"} \\
      0, & \text{"on"} 
   \end{cases}
\end{equation}
while the MF variable $n(\theta_i)=\langle s_i\rangle$ denotes the probability that MU located at $\theta_i$ is active.

\begin{figure} [htbp] 
\centerline{\includegraphics[width=1\textwidth]{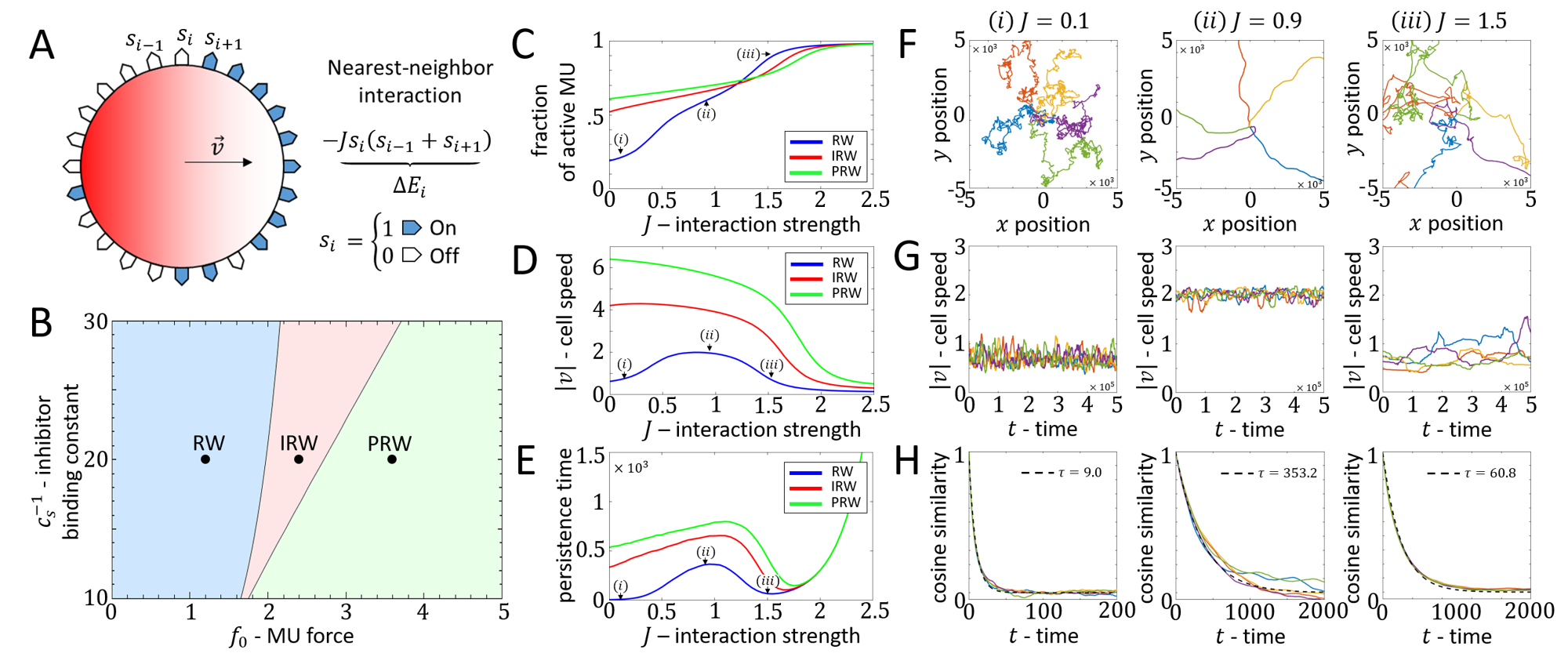}}
\caption{\footnotesize{Effect of nearest-neighbor interactions between Motile Units. A) Schematic illustration of the MU model with an additional ferromagnetic nearest-neighbor interaction of strength $J$. B) The $c_s^{-1}-f_0$ parameter space, indicating the parameter sets studied in (C–H) for $c_s^{-1}=20$ and $f_0=1.2,2.4,3.6$, corresponding to the RW, IRW, and PRW regimes, respectively. C) Fraction of active MU ($n_i=1$) as a function of the interaction strength $J$. D) Mean cell speed as a function of the interaction strength $J$. E) Directional persistence time as a function of the interaction strength $J$. The blue, red, and green curves in (C–E) correspond to the RW, IRW, and PRW regimes, respectively (parameter sets indicated in B). F) Representative cell trajectories. G) Time series of the cell speed. H) Velocity-direction cosine similarity as a function of time. Panels (i), (ii), and (iii) correspond to interaction strengths $J=0.1,0.9,1.5$, respectively, as indicated in (C–E). Other model parameters: $R=1$,$D=1$,$c_{tot}=1$.}}
\label{fig:MU_03}
\end{figure}

Due to the direct interaction of each MU with its two neighbors (Fig.\ref{fig:MU_03}A), there is an additional energy for their direct (ferromagnetic) spin-spin interaction
\begin{equation}
E_i = -Js_i\left(s_{i-1} + s_{i+1}\right)
\end{equation}
where $J$ is the magnitude of the effective NN interaction.
A spin-flip at site $i$ amounts to: $s_i\rightarrow 1-s_i$, which gives the following energy change due to the NN interaction
\begin{equation}
\Delta E_i = -J(1-2s_i)\left(s_{i-1} + s_{i+1}\right)
\end{equation}

Combined with the advection-based rates (Eqs.\ref{n_dot},\ref{ntheta}), the stochastic transition rates are now given by
\begin{eqnarray}
k_{on,i}\,&= &k_{on}\,e^{\,-Jm_i}\\
k_{off,i}&= &k_{off}\,c\left(\theta_i\right)e^{\,Jm_i}
\end{eqnarray}
where $m_i=s_{i-1} + s_{i+1}$.

Replacing the binary variables by their mean values $s_i\rightarrow n(\theta_i)$ gives the corresponding mean-field equation
\begin{equation}
\dot{n}(\theta_i) = k_{\text{on}} (1 - n(\theta_i)) e^{-Jm_i} - k_{\text{off}} n(\theta_i) c(\theta_i) e^{Jm_i}
\end{equation}
where $m_i\rightarrow n(\theta_{i-1})+n(\theta_{i+1})$.

At steady state we get
\begin{equation}
n(\theta_i) = \frac{k_{\text{on}} e^{-J m_i}}{k_{\text{on}} e^{-J m_i} + k_{\text{off}} c(\theta_i) e^{J m_i}} = \frac{k_{\text{on}}}{k_{\text{on}} + k_{\text{off}} c(\theta_i) e^{2J m_i}}
\end{equation}

The additional NN interactions have only a minor effect on the phase diagram of the migration modes, shown in Fig.\ref{fig:MU_03}B (compare to Fig.\ref{fig:MU_02}A). However, they substantially modify the migration dynamics within each migration mode, as shown in Fig.\ref{fig:MU_03}C-E. 

In the RW phase, increasing $J$ promotes the activation of the MU, and thereby the formation of a stable cluster of active MUs (Fig.\ref{fig:MU_03}C), which leads to a large increase in both the mean cell speed (Fig.\ref{fig:MU_03}D), and the directional persistence time (Fig.\ref{fig:MU_03}E). These changes are demonstrated in Fig.\ref{fig:MU_03}F-H for the values of $J$ denoted by (i)-(ii) in Fig.\ref{fig:MU_03}C-E.  
At sufficiently large values of $J$, nearly all the MU become active due to the ferromagnetic NN interactions (Fig.\ref{fig:MU_03}C), resulting in global cancellation of the protrusive forces around the cell rim and, consequently, a strong reduction in the cell speed (Fig.\ref{fig:MU_03}D). The same force-cancellation mechanism dominates the behavior in the I-RW and P-RW phases, where the cell is already spontaneously polarized by the UCSP mechanism. This reduced speed is shown for point (iii) in Fig.\ref{fig:MU_03}D,G. 

For the persistence time we find a more complex behavior. The increased activation with increasing $J$ leads to larger transverse velocity fluctuations that diminish the persistence time (point (iii) in Fig.\ref{fig:MU_03}E). However, in the limit of very large $J$ the NN interaction is so strong that it inhibits any further spin-flips, thereby the persistence time diverges.

We conclude that a short-range excitatory interaction, in addition to the long-range inhibition, can contribute to stabilize more robust migration (faster and more persistent). However if this interaction is too strong, it has the opposite effects, similar to the re-entrant RW phase in the limit of high levels of MU activation (large $k_{on}$, small $c_s^{-1}$, Fig.\ref{fig:MU_01}B(i)).


\section{Comparison of the MU-model to experimental observations}

The phase-diagram of our model (Fig.\ref{fig:MU_01}B), predicts the characteristics of cell migration as function of the model parameters $c_s^{-1}$ and $f_0$. These parameters correspond, respectively, to the inverse probability of initiating a protrusion (activation of the MU) at the cell rim, and to the magnitude of the protrusive forces exerted by the MU, which act to polarize the cell when incorporated with the USCP mechanism. From the model phase diagram, we expect cells to be non-polar and exhibit non-persistent (RW) migration for either very weak activity (low $k_{on}$, high $c_s^{-1}$), or when the entire cell rim is activated (high $k_{on}$, low $c_s^{-1}$), as indicated by the RW phase in Fig.\ref{fig:MU_01}B.

Cells with very high levels of actin polymerization activation along their periphery are observed to move but with low directionality (low persistence length $l_p$), due to the formation of new protrusions in multiple directions \cite{pankov2005rac,fort2018fam49}. This was shown when an inhibitor of actin nucleation at the cell rim (CYRI) was knocked-down \cite{fort2018fam49}. The WT cells, with normal levels of this inhibitor, exhibited higher persistence and chemotactic migration, suggesting that these cells evolved to have maximal speed and directionality. Consistent with this observation, we showed in Fig.\ref{fig:MU_01}D that the persistence length of cells in our MU model peaks at some finite level of MU activation. Furthermore, in the limit of vanishing $c_s^{-1}$ the MU model predicts that the almost uniform activation of MU along the entire cell periphery causes the P-RW regime to transition into the RW regime (Fig.\ref{fig:MU_01}B). A similar transition occurs in our model under strong local NN positive feedback between the MU (Fig.\ref{fig:MU_03}D). This is consistent with many hyper-activated cells becoming non-motile \cite{fort2018fam49}.

More generally, increased activation of actin polymerization can lead to increase in cell speed and persistence \cite{maiuri2015actin}, as we show for decreasing $c_s^{-1}$ in panels (v)-to-(ii) in Fig.\ref{fig:MU_01}D. Experimentally, the increased activation can appear when inhibitory components are reduced \cite{dang2013inhibitory,fokin2024inactivating}, and an increase in cell speed and persistence is observed.

Run-and-tumble migration, as predicted in the I-RW phase (Fig.\ref{fig:MU_02}), has been observed in many cell types \cite{li2008persistent,maiuri2015actin,diz2016steering,olguin2021chemokine}. The MU model predicts that increasing protrusive activity (decreasing $c_s^{-1}$) leads to longer run phases and shorter tumbles (Fig.\ref{fig:MU_02}F,G). This is in agreement with experimental observations \cite{gorelik2015arp2}, where lower actin inhibition correlated with these predicted changes to the run-and-tumble dynamics. In \cite{zhang2023run} a similar behvaior was observed, where faster (more active) cells exhibit longer run phases, as our model predicts.

In many amoeboid cells that perform free random migration \cite{selmeczi2008cell} the run phases are characterized by having a well defined leading-edge protrusion, whereas tumble phases have multiple smaller and competing protrusions \cite{van2017coupled}. This behavior is naturally produced by the MU model, in which the MU organize into a well-polarized leading-edge cluster during run phases, which destabilizes in the tumble phase, as shown by the dynamics of the polarity index (Fig.\ref{fig:MU_02}C). 
We note also that cells may utilize in addition internal oscillatory signals to exert fine control over the rate of tumble events \cite{hoffmann2025corrections}.

A similar distinction between persistent and random migration has been observed for bleb-driven cell migration, where persistent migration is associated with a single long-lived bleb, while random motion appears when many blebs form along the entire cell surface \cite{reichman2004autonomous,diz2016steering,waterman2026blebs}. Excessive bleb activation also reduces persistence \cite{blaser2006migration}, in agreement with our MU model, where the persistence length decreases at high $k_{on}$, and low $c_s^{-1}$ (change from point (ii) to (i) in Fig.\ref{fig:MU_01}D), until the RW phase re-appears (Fig.\ref{fig:MU_01}B).

These comparisons validate the overall features of the cell-migration properties predicted by the MU model. In the next sections we demonstrate two applications of the MU model by describing cellular chemotaxis and the response of migrating cells to localized photoactivation.

\begin{figure} [htbp] 
\centerline{\includegraphics[width=1\textwidth]{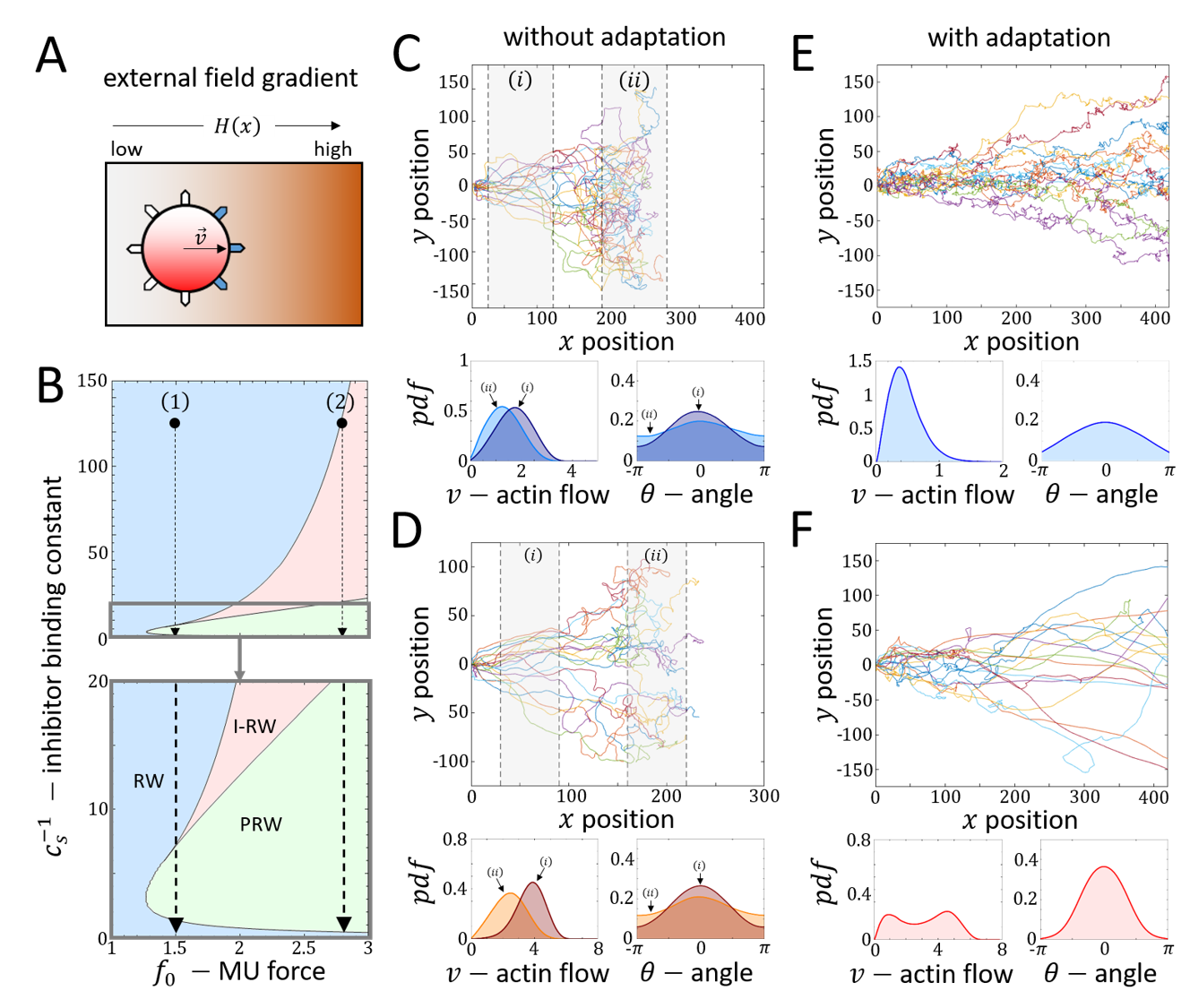}}
\caption{\footnotesize{ A) Illustration of the motile units model in the presence of an external field $H(x)$, representing the chemoattractant concentration gradient. B) The $c_s$-$f_0$ phase diagram. Point (1) refer to results shown in panels (C) and (E). Point (2) refer to results shown in panels (D) and (F). Arrows indicate the effective inhibitor binding constant as the external field increases. C-D) Upper panel: Trajectories of the model without adaptation for point (1-2). Lower panels: Velocity and angle distributions for regions (i) and (i) (gray regions above). E-F) Upper panels: Trajectories of the model with adaptation for point (1-2). Lower panels: Velocity and angle distributions of the entire time series.  Parameters: $D=1$, $R=1$, $c_{tot}=1, h_0=1,h_s=0.1$. }}
\label{fig:MU_04}
\end{figure}

\section{Chemotaxis in the MU model}

A chemoattractant affects cellular motility, through receptors at the cell membrane, which signal a local increase in the protrusive activity of the actin polymerization \cite{hecht2011activated}, as well as blebing activity \cite{tarbashevich2015chemokine,olguin2021chemokine}. This process ends up polarizing the cell towards increasing chemoattractant concentration, and chemotactic migration \cite{swaney2010eukaryotic,vaidvziulyte2022persistent}.

Within our model the cell's ability to sense an external chemoattractant is most naturally implemented as an increase in a concentration-dependent activation probability of the MU ($k_{on}$ in Eq.(\ref{n_dot})). Alternatively, or in addition, one can consider that the chemoattractant signals an increase in the strength of the MU's contribution to the retrograde flow, such that the contribution to the retrograde flow from each active MU is locally dependent on the concentration of the chemoattractant along the edges of the cell (as was used in \cite{liu2026modelling}). Both implementations are similar in spirit to the increase in pseudopod activation and persistence used to model chemotaxis \cite{alonso2025persistent}.

We introduce a concentration-dependent activation rate of the MU
\begin{equation} \label{k_on_c}
k_{on}\left(\theta_i\right)=c_s\left(1+\frac{H\left(\theta_i\right)}{h_s}\right)
\end{equation}
where $H\left(\theta_i\right)$ is the concentration of the chemoattractant at the location of MU $i$ (Fig.\ref{fig:MU_04}A). 

For a linear gradient along the $x$-axis (Fig.\ref{fig:MU_04}A), the local concentration changes along the cell rim according to 
\begin{equation}
H\left(\theta_i\right)=h_0(x+Rcos\left(\theta_i\right))
\end{equation}
where $x$ is the location of the cell's center of mass and $h_0$ is the amplitude of the concentration gradient.

As the cell moves up the gradient, it is effectively increasing its global average $k_{on}$ (Eq.(\ref{k_on_c})) and therefore decreasing $c_s^{-1}$, causing its migration state to move downward through the phase-diagram (vertical arrows in Fig.\ref{fig:MU_04}B). We consider two representative initial conditions, corresponding to cells starting at points (1) and (2) in Fig.\ref{fig:MU_04}B, respectively. 

For cells starting from point (1) in the phase diagram (Fig.\ref{fig:MU_04}B), we show typical cellular trajectories as they migrate up the gradient (Fig.\ref{fig:MU_04}C). As the cells move up the gradient, $c_s^{-1}$ is decreasing, and the cells transition from the RW to the PRW phase (denoted by (i) in Fig.\ref{fig:MU_04}C). In this regime, the trajectories become more directed and the migration speed increases, which demonstrate efficient chemotaxis. The bias toward the gradient is reflected in the speed and angular distributions in the lower panels of Fig.\ref{fig:MU_04}C. As the cells continue to migrate toward higher concentrations, they re-enter the RW phase (region (ii) in Fig.\ref{fig:MU_04}C, lower panel in Fig.\ref{fig:MU_04}B). In this regime, $k_{on}$ is high and most MU are active, which leads to stronger cancellation of the net force, lower migration speed, and a weaker chemotactic bias. The same qualitative trends, but more pronounced. are observed for cells starting from point (2), as shown in Fig.\ref{fig:MU_04}D.

This chemotactic behavior, whereby the directional migration is maximal at intermediate chemoattractant concentrations but decreases at very high concentrations, agrees with previous models based on explicit saturation of the local receptors response to the chemoattractant \cite{malet2015collective}. Lower chemotactic migration in higher background levels of the chamoattractant also appear in a pseudopod model of chemotaxis \cite{alonso2025persistent}, and models based on receptor saturation \cite{rappel2008receptor,ferguson2017statistical,eidi2017modelling} and on signal-to-noise ratio calculations \cite{ueda2007stochastic}. In our model, this saturation effect emerges implicitly from the loss of cell polarity at very high global activation. Similar behavior, of lower chemotactic directional migration for higher background concentrations, was previously observed experimentally for chemotaxis of small cellular clusters \cite{cai2016modeling}, and for isolated cells \cite{herzmark2007bound,ueda2007stochastic,eidi2017modelling,panigrahi2026chemotrack}.

In order to explore the inherent response of the cells in the different phases to the chemical gradient, we also considered a concentration gradient that does not change with the spatial location of the cell, in the form of
\begin{equation} \label{H2}
H\left(\theta_i\right)=h_0\left(\frac{1+cos\left(\theta_i\right)}{2}\right)
\end{equation}
Beyond serving as a computational way to study long trajectories within the same chemical gradient \cite{sanoria2026chemotaxis}, this situation corresponds to "perfect adaptivity", if a cell could distinguish concentration differences irrespective of the absolute background concentration \cite{adler2018fold}.

The results of the adaptive model (Eq.(\ref{H2})) are shown in Fig.\ref{fig:MU_04}E for cells starting in the RW phase (point (1) in Fig.\ref{fig:MU_04}B), and in Fig.\ref{fig:MU_04}F for cells starting in the I-RW phase (point (2) in Fig.\ref{fig:MU_04}B). The chemotactic response is found to be stronger in the I-RW phase, where cells spontaneously polarize in the direction of the chemical gradient (Fig.\ref{fig:MU_04}E). In contrast, cells starting in the RW phase exhibit weaker and more transient polarization, resulting in less pronounced chemotactic response (Fig.\ref{fig:MU_04}F).

\section{Localized photoactivation in the MU model}

As a final demonstration of the utility of the MU model, we explore its response to localized external activation. We are motivated by optogenetic photoactivation experiments \cite{vaidvziulyte2022persistent,town2023local}, in which a localized laser beam enhances the activity of actin polymerization at specific locations along the cell periphery. 

\begin{figure} [htbp] 
\centerline{\includegraphics[width=1\textwidth]{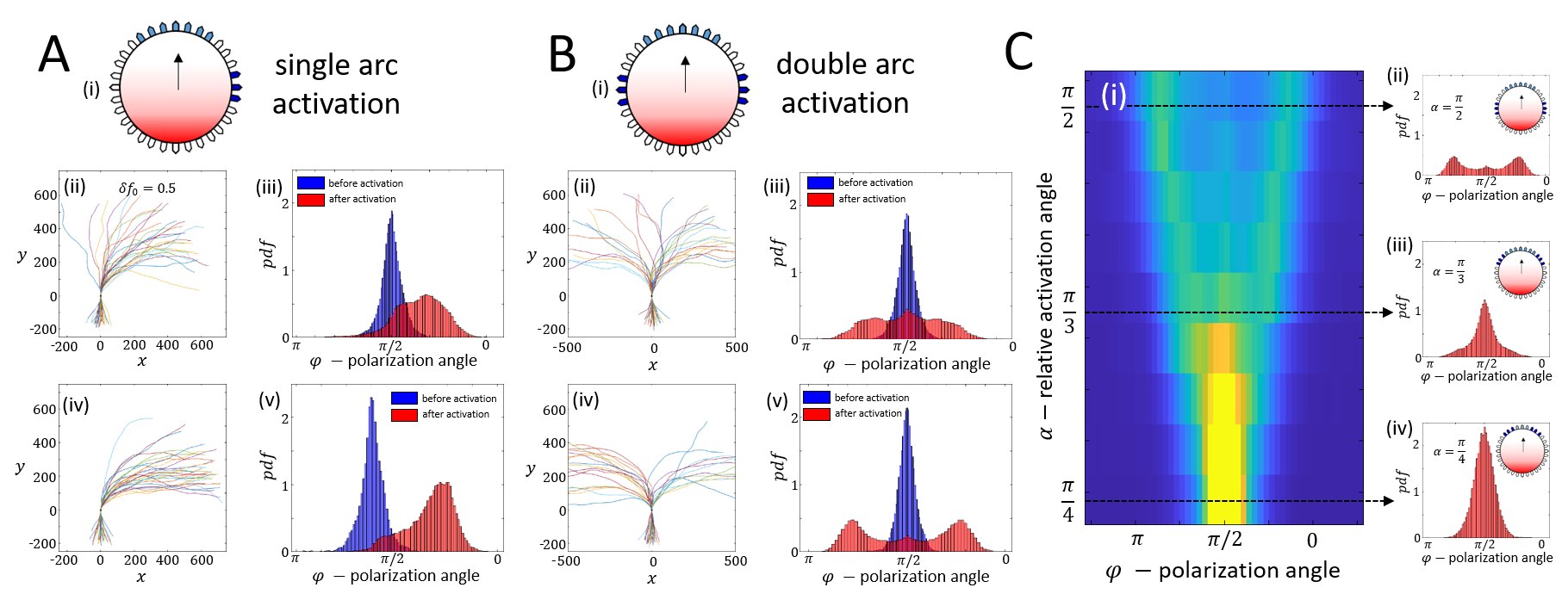}}
\caption{\footnotesize{ A) Single arc activation at angle $\alpha=\pi/2$ with respect to the initial polarity direction. The cell is in the PRW regime ($c_s^{-1}=10,f_0=2.5$, see Fig.\ref{fig:MU_01}B(i)), for a system with $N=64$ MUs. i) Illustration of the activation model. ii-iii) Cell trajectories and polarization angle distribution for $\delta f_0=0.5$. iv-v) Cell trajectories and polarization angle distribution for $\delta f_0=1$.  Red/blue histograms indicate trajectories before/after activation (500 trajectories). B) Symmetric double arcs activation, at angle $\alpha=\pm\pi/2$ with respect to the initial polarity direction. i) Illustration of the activation model. ii-iii) Cell trajectories and polarization angle distribution for $\delta f_0=0.5$. iv-v) Cell trajectories and polarization angle distribution for $\delta f_0=1$.  Red/blue histograms indicate trajectories before/after activation (500 trajectories). C) Panel (i) shows the heatmap of the polarization angle with respect to the relative activation angle of two symmetric patches with respect to the initial polarity direction, for $\delta f_0=1$. Panels (ii,iii,iv) display the polarization angle distribution for $\alpha=\pi/2,\pi/3,\pi/4$. Other parameters: $D=1$, $R=1$, $c_{tot}=1.$}}
\label{fig:MU_05}
\end{figure}

To better resolve the spatial extent of the photoactivated region, we increased the number of MU from 32 to 64 in these simulations. We then modeled the illuminated region as an arc of three adjacent MU, a fraction of 3/64 of the cell perimeter, which approximately matches the fraction of the cell perimeter illuminated in the experiments \cite{town2023local} (Fig.\ref{fig:MU_05}A(i)).

We describe the local enhancement of actin polymerization activity due to photoactivation by increasing the value of the parameter $f_0$ in these MU, which determines the magnitude of their contribution to the net protrusive force and the retrograde flow that determines the profile of the inhibitory polarity cue (Eq.(\ref{vss2D})) 
\begin{equation}
f_0 = \bar{f}_0+\delta f_0
\label{deltaf}
\end{equation}

In Fig.\ref{fig:MU_05}A(i) we demonstrate the effect of applying photoactivation at an angle of $\alpha=\pi/2$ with respect to the initial direction of cell migration. The initial direction of migration in the lab frame is along the $y$-axis ($\varphi=\pi/2$), as shown by the blue distribution in Fig.\ref{fig:MU_05}A(iii,v). Since the cells in the experiments were highly motile and persistent, we chose a MU-cell in the PRW regime. As in the experiments \cite{town2023local}, we find that this photoactivation causes the cell to turn and migrate towards the activated patch on the rim (towards the $x$-axis, $\varphi=0$), as shown by the trajectories (Fig.\ref{fig:MU_05}A(ii,iv)) and the angular distributions at the end of these trajectories (red distributions in Fig.\ref{fig:MU_05}A(iii,v)). The turning response of the cells towards the direction of the activated arc, becomes stronger as $\delta f_0$ increases (Eq.(\ref{deltaf})).

In \cite{town2023local} it was found that when a cell is simultaneously activated by two laser beams at a relative activation angle of $\alpha=\pm\pi/2$ with respect to the initial polarity direction (as in Fig.\ref{fig:MU_05}B(i)), the cell eventually turns towards either the right or the left activation spots. The same behavior is observed in our simulations (Fig.\ref{fig:MU_05}B(iii-v)), arising from the long-range inhibition between MU on opposite sides of the cell through the advection of the inhibitor field (UCSP mechanism). Some cells take longer to undergo the spontaneous symmetry breaking, and therefore continue to migrate along their original direction, with the effects from the two opposing enhanced MU arcs effectively canceling each other. Such trajectories were also observed in the experiments \cite{town2023local}.

We further explore the directional symmetry breaking in response to two photoactivated regions along the rim, placed symmetrically on either side of the original direction of motion ($y$-axis). In Fig.\ref{fig:MU_05}C we plot the distribution of migration direction angles of the simulated cells (at the final time of the simulation) as a function of the relative activation angle $\alpha$ of the activated MU arcs with respect to the initial polarity axis. We find that when the angle between the two photoactivated regions is smaller than $2\alpha\sim2\pi/3$ degrees, the cells maintain their initial migration direction, which also corresponds to the vector sum of the activation directions. Above this critical angle, the cells exhibit a clear bifurcation, with cells migrating in the direction of either  one of the photoactivated regions. In the language of directional decision-making (DDM), moving along the vector sum of the inputs is called the "compromise" direction, while following one of the inputs is associated with a directional "decision". Our model predicts cellular DDM which resembles what was observed for DDM in animal behavior \cite{pinkoviezky2018collective,sridhar2021geometry,oscar2023simple}.

In \cite{town2023local} the photoactivation at $\alpha=\pi/4$ was tested, where about half of the cells were observed to follow the photoactivated regions, while half of the cells did not exhibit a clear bifurcation. This observed spread in cellular behavior is reminiscent to the observed spread in the bifurcations for animal trajectories during DDM \cite{sridhar2021geometry}. In our model the critical angle for the "decision" bifurcation depends on the location in the MU model phase diagram (Fig.\ref{fig:MU_01}). A systematic study of the critical angle for cells, as we show theoretically in Fig.\ref{fig:MU_05}C, awaits future experimental and theoretical studies.

\section{Discussion}

We have presented here a two-dimensional theoretical model of cell migration, where the cell spontaneously polarizes through the UCSP mechanism \cite{maiuri2015actin,ron2020one}, namely the long-range advection of an inhibitory polarity cue. We map this mechanism to a system of interacting motile-units (MU) along the cell rim, thereby discretizing the model and mapping it to a system of interacting "spins". This model naturally allows to describe the inherent stochasticity related to the process of excitation or inhibition of these MU, in terms of their transition rates. Additional interactions, beyond the UCSP mechanism, are easy to employ in the language of additional terms in the Hamiltonian of the interacting spins. The model therefore goes beyond the current one-dimensional manifestations of the UCSP model \cite{maiuri2015actin,lavi2016deterministic,ron2020one,papaganti2026minimal}, and provides a framework for describing cell migration in terms of an interacting and stochastic spin system. 

We demonstrate that this model allows for a very intuitive description of the different cellular migration modes, as well as cellular dynamics during chemotaxis. Finally, we showed that this model can explain the effects of photoactivation experiments \cite{town2023local}, and exhibits the transition from "compromise" to "decision" regarding migration direction, which was previously observed in animal navigation \cite{pinkoviezky2018collective,sridhar2021geometry,oscar2023simple}.
The model can be extended to describe the interactions between cells and external barriers, as well as other cells. Upon collision with an external barrier, we may simply turn off the MU that are facing the obstacle, which would lead to the reorientation of the polarization axis and the scattering of the cell from its original path. This could be explored in the context of experimental studies \cite{paksa2016repulsive,gross2020using,shubhadeep2025PhysRevResearch}. Similarly, our MU model can be used to explore the protrusion-based interactions of cells with complex environments, such as explored in \cite{caballero2014protrusion}.


In the same spirit, the model can provide a platform for studying interactions between cells and their effect on the emergence of collective cell migration. We can add different interactions that affect the activity of MU on neighboring cells that are within a contact distance from each other. This interaction can be symmetric and inhibitory, as in the "contact inhibition of locomotion" (CIL), which plays a major role in collective cell migration \cite{camley2014polarity,szabo2016modelling}. It can depend on the activity state of the MU, such that an active MU in contact with non-active MU on the neighboring cell gets enhanced, as in the process of "cryptic lamellipodia", thereby stabilizing cells to migrate in the same direction \cite{jain2020role}. These interactions were implemented previously in a one-dimensional version of the UCSP model \cite{ron2023polarization}, and the MU model allows to study their effects on collective cell migration in two-dimensions.

We note that a future direction could be to map the spin model to an effective continuum dynamics model of the retrograde flow $v$ as an internal dynamical degree of freedom. This would amount to treating each particle as an active Brownian particle (ABP), where its motion is driven by the "hidden" dynamics of $v$, which will be affected by interactions between the particles \cite{romanczuk2012active}. A few ABP models have included an internal dynamical degree of freedom \cite{tilch1999directed,zhang2008active}. However this direction has not been extensively explored, and our model may motivate the development of this formalism. 

Our MU model therefore offers a platform for studying an active matter system, where particles have an internal spin-ring structure. In this way we expand the scope of current active-matter models with internal degrees of freedom \cite{demaerel2018active,bebon2026thermodynamics,valani2026wave}, by introducing a model where the polarity and speed of the self-propelled particles is determined by their internal spin dynamics. We demonstrate that such particles can exhibit directional-decision making which is a rudimentary form of "deciding" active particles \cite{gompper20252025}, and can serve as the basis for more "intelligent" active particles \cite{iyer2026emergent}. This mixture of self-propulsion and spin dynamics can open new directions for the study of non-equilibrium physics.
\section{Acknowledgments}
We would like to thank Michael Assaf and Orion Weiner for discussions.
\newline

\appendix
\setcounter{equation}{0}
\renewcommand{\theequation}{A.\arabic{equation}}
\setcounter{figure}{0}
\renewcommand{\thefigure}{A.\arabic{figure}}
\section*{Appendix A: Derivation of the inhibitor concentration profile}
As in the one-dimensional UCSP model \cite{maiuri2015actin}, we assume that the polarity cue is transported by advection and diffusion along the axis of polarization which we define as the $x$-axis.
The concentration of the polarity cue therefore obeys the one-dimensional advection-diffusion equation
\begin{equation}
\frac{\partial c(x,t)}{\partial t}=D\frac{\partial^2 c(x,t)}{\partial x^2}+v\frac{\partial c(x,t)}{\partial x}
\end{equation}
where $D$ is the diffusion constant, and $v$ is the magnitude of the retrograde flow.

Next, we assume a limit of fast exchange where the polarity cue rapidly reaches a steady-state ($\partial c/\partial t =0$) and that there are no fluxes at the boundaries. Under these conditions the steady-state solution is given by
\begin{equation}
c(x) = c_0\, exp\left(-\frac{vx}{D}\right)
\end{equation}
where $c_0$ is the constant of integration.

The retrograde flow is generated collectively by all the active motile units. Each active MU contributes a unit of retrograde flow directed radially inward (Fig.\ref{fig:MU_01}A), such that the total retrograde flow is given by Eq.(\ref{vret_discrete}.)



At the cell rim ($r=R$) we obtain

\begin{equation}
c(\theta) = c_0\, exp\left(-\frac{vR}{D} cos(\theta)\right)
\end{equation}

The constant of integration $c_0$ is found using the constraint that the total amount of inhibitor in the cell is conserved. By integrating $c(\theta)$ over the total circular area of the cell we obtain
\begin{equation} 
c_{tot}=c_0\int_{0}^{2\pi}d\theta\int_{0}^R r dr \exp\left(-\frac{v r cos(\theta)}{D}\right) = c_0\frac{2D\pi R}{v} I_1\left(\frac{v R}{D}\right)\rightarrow c_0=\frac{1}{2\pi R}\frac{c_{tot} v}{D} \frac{1}{I_1\left(\frac{v R}{D}\right)}
\end{equation}

Therefore, the concentration profile of the polarity cue as a function of a MU location with respect to the polarization axis is given by
\begin{equation} \label{c2D}
     c(\theta) = \frac{c_{tot} v}{2\pi R D}\frac{exp\left(-\frac{v R}{D}cos(\theta)\right)}{I_1\left(\frac{v R}{D}\right)}
\end{equation}
where $I_1$ is the modified Bessel function of the first kind.
\newline

\section*{Appendix B: Finite size effects}

Comparisons between the simulations and the mean-field (MF) solution (Eq.(\ref{v_ss})) for the predicted phase-diagram shown in Fig.\ref{fig:MU_01}B, show that there are strong finite size effects in the simulations. For example, in Fig.\ref{fig:MU_01}B points (iii,iv) are in the RW phase, where only a zero-velocity MF solution exits. Nevertheless, a finite velocity peak is found for these points for $N=32$ MU (Fig.\ref{fig:MU_01}C(iii,iv)). 

\begin{figure} [htbp] 
\centerline{\includegraphics[width=1\textwidth]{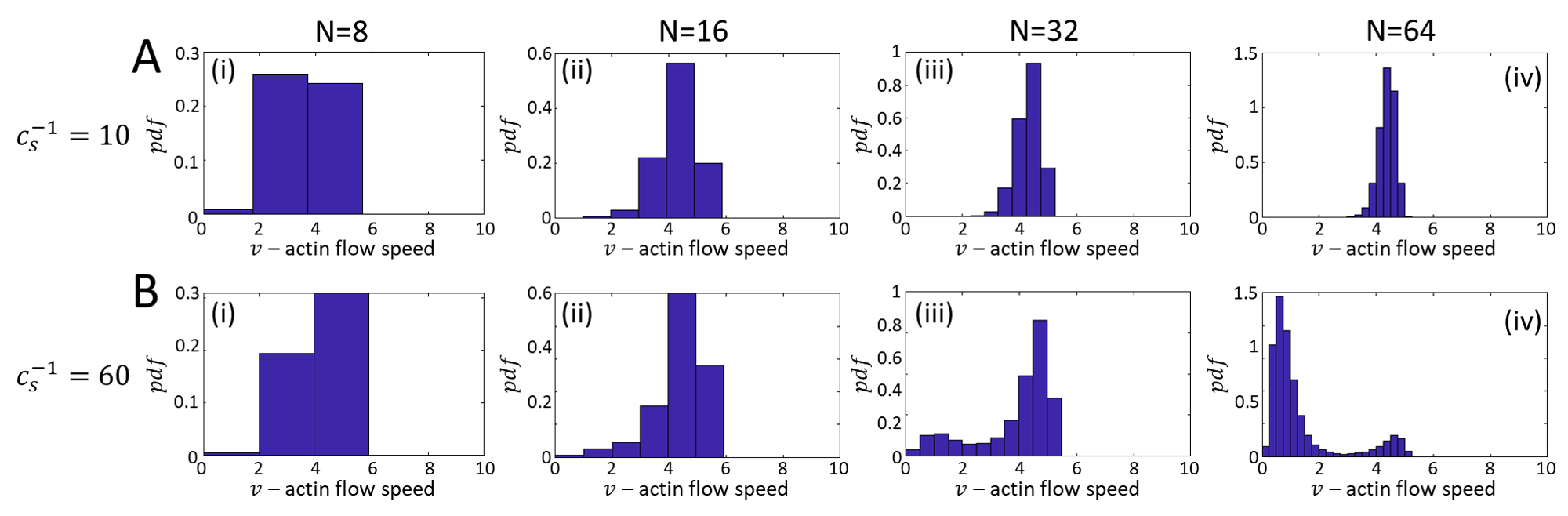}}
\caption{\footnotesize{ Histograms of the steady-state flow speed $v$ for different system size $N$ at $f_0=2.5$ and A) $c_s^{-1}=10$ and B) $c_s^{-1}=60$. Panels (i-iv) correspond to $N=8,16,32$ and $64$ respectively. }}
\label{fig:MU_finite_size}
\end{figure}

In Fig.\ref{fig:MU_finite_size} we further explore the effects of the number $N$ of MU. In Fig.\ref{fig:MU_finite_size}A the parameters correspond to point (i) in Fig.\ref{fig:MU_01}B, which is in the PRW regime according to the MF phase diagram. As $N$ increases the velocity distribution becomes more sharply peaked around the MF solution value, yet the velocity distribution remains with a single finite-value peak as expected for the PRW regime also for small values of $N$.

In Fig.\ref{fig:MU_finite_size}A the parameters correspond to point (ii) in Fig.\ref{fig:MU_01}B, which is in the I-RW regime according to the MF phase diagram. We clearly observe that as $N$ decreases the velocity distribution becomes more similar to the PRW phase with a single finite-velocity peak. Only at larger $N=32,64$ the double-peak distribution of the I-RW is recovered. Within the double-peak distribution we find that the relative weight of the peaks also changes with increasing $N$. Since point (ii) is very close to the transition from I-RW to RW, we expect that the average cell speed predicted by the MF solution will very small, as shown by the $N=64$ system. The large $N$ limit more closey matches the MF behavior, as expected.

\section*{Appendix C: Persistence length calculation}
The persistence time and persistence length can be estimated from the fluctuations in the cell velocity caused by the spin-flipping dynamics of the MUs (Eq.(\ref{n_dot})).

We assume that the cell is polarized along the $x$-axis, such that the mean velocity in the perpendicular direction is zero.
A spin-flip of the $i$-th MU changes the velocity in the perpendicular direction by
\begin{equation}
\delta v_{\perp,i}=\frac{2\pi f_0}{N} sin(\theta_i)
\end{equation}
The probability of an MU to turn on (off) at a short time interval $\delta t$ is $k_{on}(1-n_i)\delta t$ ($k_{off}n_ic(\theta_i)\delta t$), respectively.

The variance of the perpendicular velocity is given by
\begin{eqnarray}
\sigma_{v_{\perp}}&=&\underbrace{\text{lim}}_{\delta t\rightarrow 0}\frac{ \langle\Delta  v_{\perp}^2\rangle}{\delta t} \\
&=& \left(\frac{2\pi f_0}{N}\right)^2 \sum_{i=1}^N sin^2(\theta_i)\left[k_{on}(1-n_i)+k_{off}n(\theta_i)c(\theta_i)\right] \label{var}
\end{eqnarray} 

At steady state we obtain
\begin{equation}
k_{on}(1-n(\theta_i))=k_{off} n(\theta_i)c(\theta_i)
\end{equation}
and therefore Eq.(\ref{var}) becomes
\begin{eqnarray}
\sigma_{v_{\perp}}=2\left(\frac{2\pi f_0}{N}\right)^2 k_{off}\sum_{i=1}^N sin^2(\theta_i)n(\theta_i)c(\theta_i)
\end{eqnarray}
and using the steady state probability of an MU to be active (Eq.(\ref{n_dot})) we obtain
\begin{eqnarray}
\sigma_{v_{\perp}}(v)=2\left(\frac{2\pi f_0}{N}\right)^2 k_{off}\sum_{i=1}^N\frac{sin^2(\theta_i)c(\theta_i,v)}{1+c(\theta_i,v)/c_s}
\end{eqnarray}

The discrete sum is retained here since the fluctuations originate from the finite number of MUs. The prefactor scales as $N^{-2}$, whereas the sum containes $N$ terms, and therefore the transverse fluctuations scale approximately as $N^{-1}$, and vanish in the mean-field model ($N\rightarrow\infty$).

We next relate the transverse velocity fluctuations to the persistence time. For a cell moving with velocity $v$ along its polarization axis, a small transverse fluctuation changes the direction of the velocity by
\begin{equation} \label{dphi}
\Delta\varphi\simeq \frac{\Delta v_{\perp}}{v}
\end{equation}
By combining Eqs. (\ref{var}) and (\ref{dphi}) we obtain that
\begin{equation}
\frac{\sigma_{v_{\perp}}}{v^2}=\frac{ \langle\Delta  \varphi ^2\rangle}{\delta t}
\end{equation}

The directional velocity correlation is defined as
\begin{equation} \label{C(tau)}
C(\tau)\equiv\langle\hat{v}(t)\cdot\hat{v}(t+\tau)\rangle=\langle cos(\varphi(t+\tau)-\varphi(t)\rangle
\end{equation}

For Gaussian angular fluctuations

\begin{equation} \label{cos(phi)}
\langle cos(\Delta \varphi)\rangle = exp\left(-\frac{1}{2}\langle\Delta \varphi^2\rangle\right)
\end{equation}

By combining Eqs. (\ref{C(tau)}) and (\ref{cos(phi)}) we obtain
\begin{equation}
C(\tau)\simeq exp\left(-\frac{\sigma_{v_{\perp}}}{2v^2}\tau\right)
\end{equation}
Since the persistence time $\tau_p$ is defined by
\begin{equation} \label{C(tau2)}
C(\tau)=exp\left(-\frac{\tau}{\tau_p}\right)
\end{equation}
we obtain
\begin{equation} \label{tp}
\tau_p=\simeq \frac{2v^2}{\sigma_{v_{\perp}}}
\end{equation}
and the persistence length $l_p$ is therefore
\begin{equation} \label{lp}
l_p=v\tau_p\simeq\frac{2v^3}{\sigma_{v_\perp}} 
\end{equation}

\begin{figure} [htbp] 
\centerline{\includegraphics[width=0.75\textwidth]{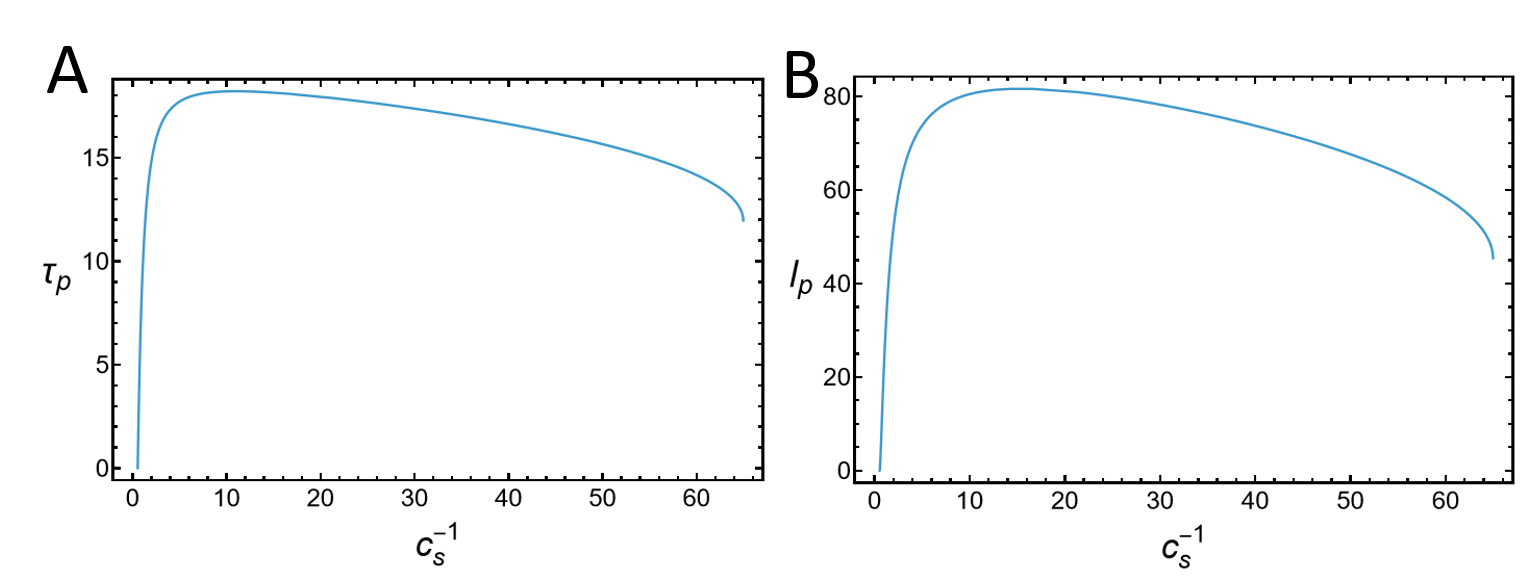}}
\caption{\footnotesize{ (A) Persistence time $\tau_p$ calculated from Eq.(\ref{tp}) along the stable nonzero mean-field solution for \(f_0=2.5\) and \(N=32\). (B) Corresponding persistence length from Eq.(\ref{lp}).}}
\label{fig:MU_app_C}
\end{figure}

Fig.\ref{fig:MU_app_C} shows $\tau_p$ and $l_p$ calculated from Eqs.\ref{tp},\ref{lp} using the stable non-zero mean-field solution at $f_0=2.5$ (corresponding to value studied in Fig. \ref{fig:MU_01}B). Both quantities show a non-monotonic dependence on $c_s^{-1}$: first an increase in the persistence followed by a decrease after reaching a maximum at intermediate values of $c_s^{-1}$, which is qualitatively consistent with the trend observed in the simulations (Fig.\ref{fig:MU_01}F).

\begin{figure} [htbp]
\centerline{\includegraphics[width=1\textwidth]{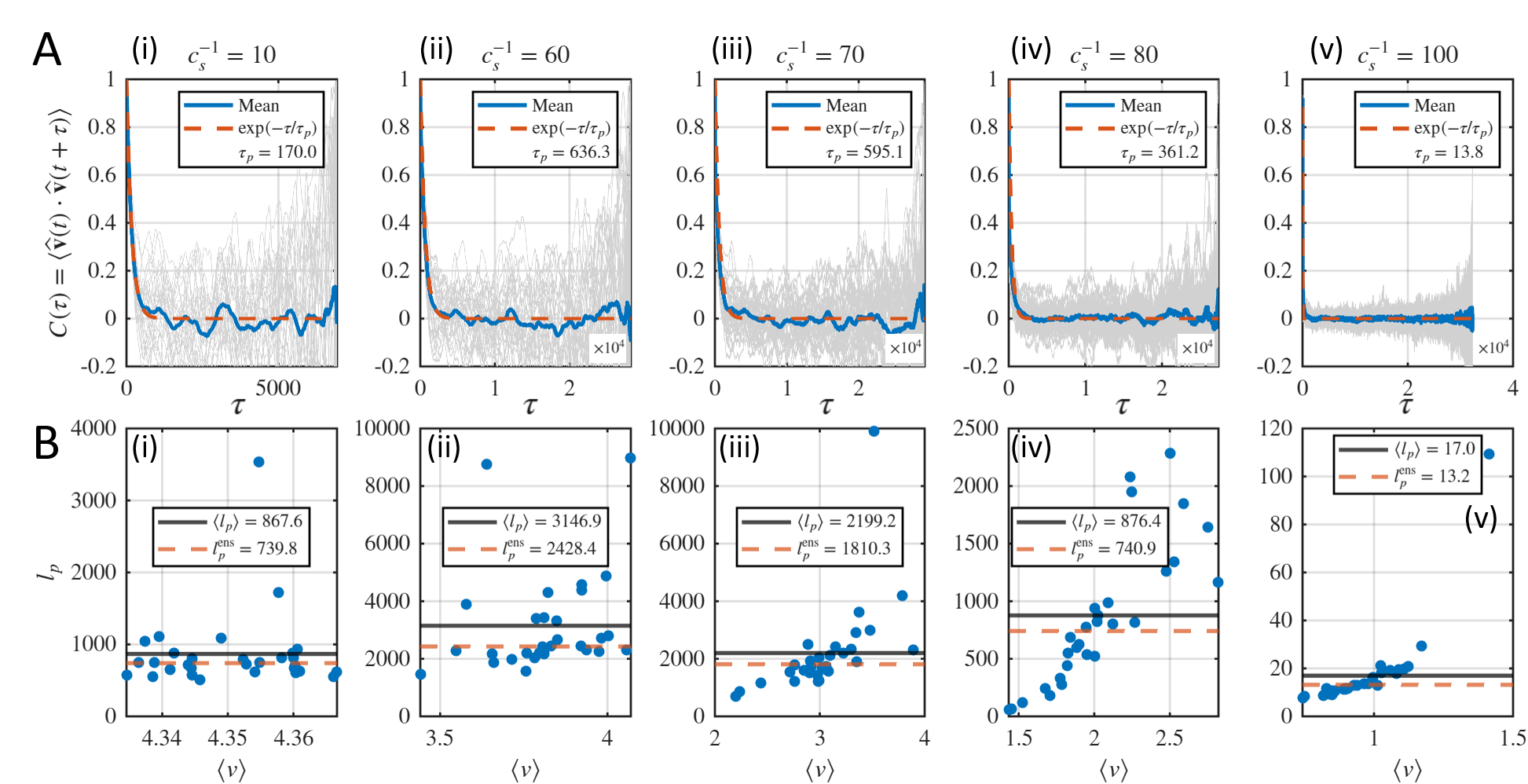}}
\caption{\footnotesize{(A) Directional velocity correlation for the points shown in Fig.\ref{fig:MU_01}(B-F). Blue curves show the mean correlation and dashed red curves show the exponential fit. The resulting $\tau_p$ is indicated in each panel. (B) Persistence lengths of the individual stochastic trajectories as a function of their mean velocity. Solid horizontal lines denote the mean persistence length $\langle l_p\rangle$ reported in the main text (Fig.\ref{fig:MU_01}F), and dashed lines denote the ensemble estimate $l_p^{\rm ens}$.}}
\label{fig:MU_app_D}
\end{figure}

To calculate the persistence time numerically for the simulated trajectories (Fig. \ref{fig:MU_01}F), we calculate the directional velocity correlation for each stochastic trajectory, where the initial decay of the correlation is fitted to Eq.(\ref{C(tau2)}), from which $\tau_p$ is obtained.
The persistence length of each trajectory is then calculated as $l_p=\langle v\rangle \tau_p$.
The persistence time reported in Fig. \ref{fig:MU_01}F in the main text is the mean of the persistence lengths observed independently for the difference trajectories
\begin{equation}
    \langle l_p \rangle = \frac{1}{N_{traj}}\sum_j\langle v \rangle_j \tau_{p,j}
\end{equation}
For comparison, we also averaged the directional correlation functions over all trajectories and fit the resulting mean correlation to $exp(-\tau/\tau_p^{ens})$, whereby the corresponding ensemble persistence length is $l_p^{ens}=\langle v \rangle_{ens}\tau_p^{ens}$.
Figure \ref{fig:MU_app_D} shows the numerical calculations for $c_s^{-1}=10,60,70,80,100$ which correspond to Fig.\ref{fig:MU_01}.

\section*{Appendix D: Polarity index}

\begin{figure} [htbp] 
\centerline{\includegraphics[width=1\textwidth]{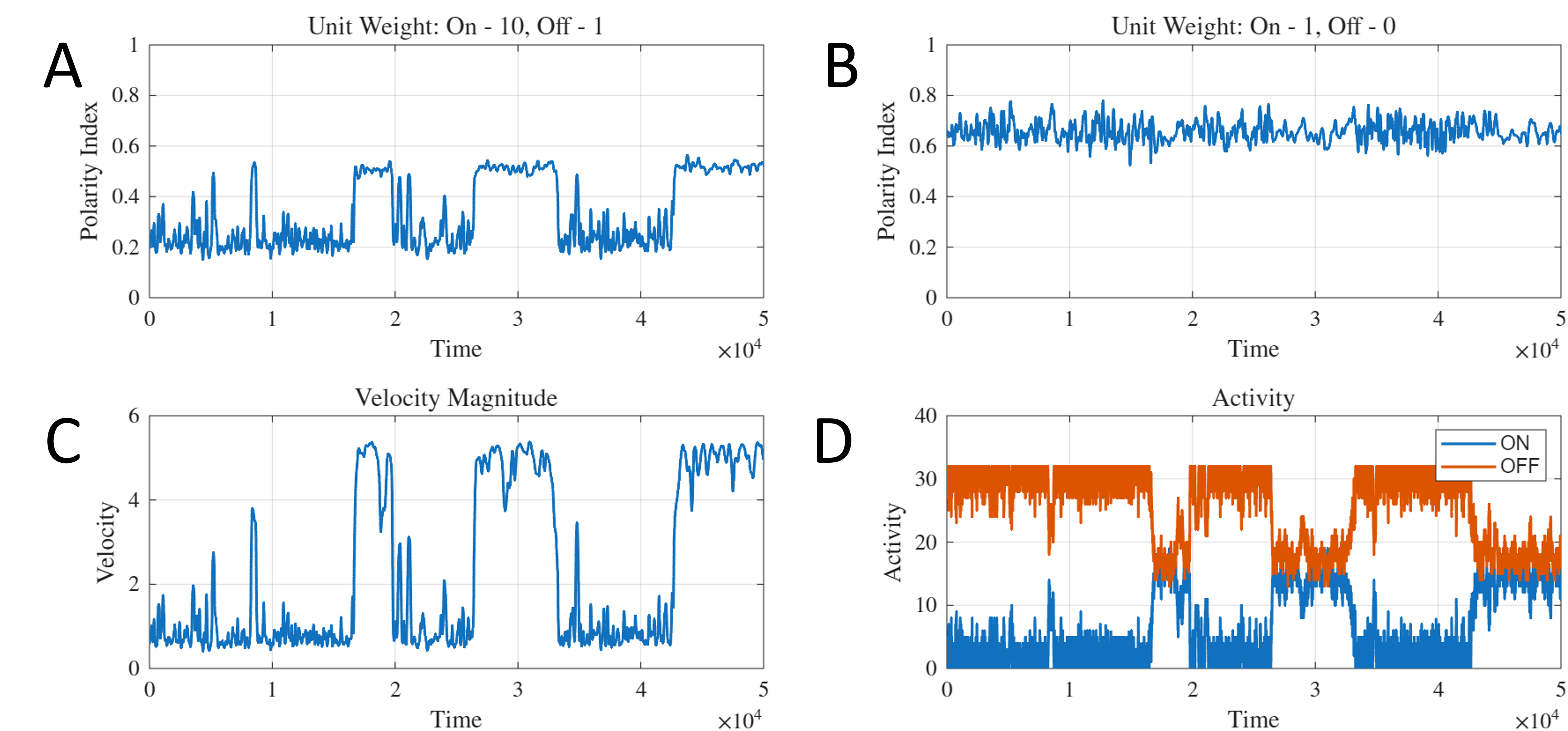}}
\caption{\footnotesize{Polarity index for a trajectory at point $(c_s^{-1},f_0)=(125,2.8)$. A) The polarity index as a function of time with unit weights of $|n_{on}|=10$ and $|n_{off}|=1$, as in Eq.(\ref{polarity}) B) The polarity index as a function of time with unit weights of $|n_{on}|=1$ and $|n_{off}|=0$, as in Eq.(\ref{polarity1}). C) The velocity magnitude ($\sqrt{v_x^2+v_y^2}$ ) as a function of time. D) The number of active MU as function of time (On - Blue, Off - Orange).}}
\label{fig:MU_app_B}
\end{figure}

We wish to motivate the choice of the polarity index as defined in Eq.(\ref{polarity}). In Fig.\ref{fig:MU_app_B} we show the time-series of a typical trajectory of a cell in the IRW phase, which undergoes run-and-tumble motility. In Fig.\ref{fig:MU_app_B}A we show the polarity index of this cell given by Eq.(\ref{polarity}), which clearly allows us to distinguish the run and tumble phases correctly, as reflected by the velocity magnitude of the cell in Fig.\ref{fig:MU_app_B}C. 

If we chose the weight of the MU sinmply according to their activity, such that $|n_{on}|=1$ and $|n_{off}|=0$, the polarity index would be defined as
\begin{equation}
p=\frac{1}{N}\sum\limits_{i=1}^{N}|n_i(\theta_i)|cos(\alpha)\label{polarity1}
\end{equation}
and this would result in the time-series shown in Fig.\ref{fig:MU_app_B}B. As can be seen, the run and tumble phases are very hard to distinguish. 

The reason for this behavior becomes clear when we plot the corresponding dynamics of the total number of activated MU in Fig.\ref{fig:MU_app_B}D. We find that during the tumble phases there are less active MU distributed homogeneously around the cell, resulting in strong cancellation and low speed. Nevertheless, their polarity index according to Eq.(\ref{polarity1}) is very noisy but can be dominated by a few MU. During the run phases there are more activated MU, and they are more polarized. Overall the change between the two phases is indistinguishable using this definition, and we therefore found the definition of Eq.(\ref{polarity}) much more useful to clearly distinguish between the run and tumble phases.

\newpage

\bibliographystyle{apsrev4-1}
\bibliography{bibliography}

\end{document}